\documentclass[preprint,showpacs,preprintnumbers,amsmath,amssymb,nofootinbib]{revtex4-1}

\usepackage{graphicx}
\usepackage{dcolumn}
\usepackage{amssymb}
\usepackage{mathrsfs}
\usepackage{amsmath}
\usepackage{epsfig}
\usepackage{hhline}
\usepackage[utf8]{inputenc}
\usepackage{color}
\usepackage{multirow}
\usepackage[T1]{fontenc}
\usepackage{verbatim}
\usepackage{slashed}
\usepackage{epstopdf}
\usepackage{lineno,hyperref}
\usepackage{cancel}
\usepackage{xcolor}
\usepackage{fancyhdr}

\begin{document}

\title{Probing single stop production at the FCC-hh/SPPC}

\author{Tian-Peng Tang}
\author{Hang Zhou}
\email[Corresponding author:]{ zhouhang@njnu.edu.cn}
\author{Ning Liu}
\email[Corresponding author:]{ liuning@njnu.edu.cn}

\affiliation{Department of Physics and Institute of Theoretical Physics, Nanjing Normal University, Nanjing, 210023, China}

\begin{abstract}
Top squark (stop) is a crucial part of supersymmetric models (SUSY) to understand the naturalness problem. Other than the traditional stop pair production, the single production via electroweak interaction provides signals with distinctive features which could help confirm the existence of the top squark. In this paper, we investigate the observability of stop through the mono-top channel of the single stop production at the future proton-proton colliders, FCC-hh and SPPC, in a simplified Minimal Supersymmetric Standard Model (MSSM). With the integrated luminosity of 3000 $\text{fb}^{-1}$, we can probe the stop with mass up to 3.25 TeV by the mono-top channel at $5\sigma$ level. Considering the systematic uncertainty of 10\%, the exclusion limit for stop mass can be reached at about 1.5 TeV. Exclusion limits on stop mass and higgsino mass parameter $\mu$ are also presented.
\end{abstract}
\maketitle

\section{Introduction}
The search for new physics beyond the Standard Model (BSM) is a primary goal for current and future colliders, though the Standard Model (SM) has been a huge success. One of the main motivations for BSM is the hierarchy problem caused by the Higgs mass quadratic divergence. Especially after the SM-like Higgs boson was discovered by the ATLAS~\cite{higgs-atlas} and CMS~\cite{higgs-cms} collaborations in 2012, new physics is expected to appear at the TeV scale to stabilize the Higgs mass without fine-tuning. The low energy supersymmetry (SUSY) is one of the most appealing and natural BSM models that can solve the hierarchy problem, by introducing superpartners of SM particles and imposing supersymmetry between fermions and bosons.

Among the supersymmetric particles predicted by SUSY, the scalar top quark (stop), which is the SM top quark's superpartner, can protect the Higgs mass by canceling out the quadratic divergence of the top quark loop, and thus, serve as an elegant solution to the hierarchy problem. Therefore, searching for the stop has always been crucial to test SUSY naturalness~\cite{nsusy-1,nsusy-2,nsusy-3,nsusy-4,nsusy-7,nsusy-9,nsusy-12,Backovic:2015rwa,dutta,nsusy-14,nsusy-15,Goncalves:2016tft,Goncalves:2016nil,Wu:2018xiz,Abdughani:2018wrw,Duan:2017zar,Han:2016xet,Arkani-Hamed:2015vfh,Tang:2019nyp,Duan:2019ykd,Zhou:2019alr,vanBeekveld:2019tqp,DiazCruz:2001gf,Kobakhidze:2015scd}. During the LHC Run-1 and Run-2, the stop has been searched for through the gluino-mediated single production and pair production by the ATLAS and CMS collaborations. The search strategies depend on a variety of kinematically allowed phase spaces of the stop decay, which can be defined by the mass-splitting $\Delta m = m_{\tilde t_1} - m_{\tilde{\chi}^{0}_{1}}$. When $\Delta m$ is much larger than the top quark mass, the top quark from stop decay is energetic. By using endpoint observables~\cite{Lester:1999tx, Bai:2012gs, Cao:2012rz} and boosted techniques~\cite{Plehn:2012pr,Plehn:2011tf,Plehn:2010st}, the stop signal can be well separated from the SM $t\bar t$ background. But when $\Delta m$ approximates zero, the decay products of stop are too soft to be observed, and thus the initial- or final-state radiation jet can be used to trigger the signal events selection~\cite{Hagiwara:2013tva,Goncalves:2014axa,Cacciapaglia:2018rqf}.

\begin{figure}[t]
\centering
\includegraphics[scale=0.4]{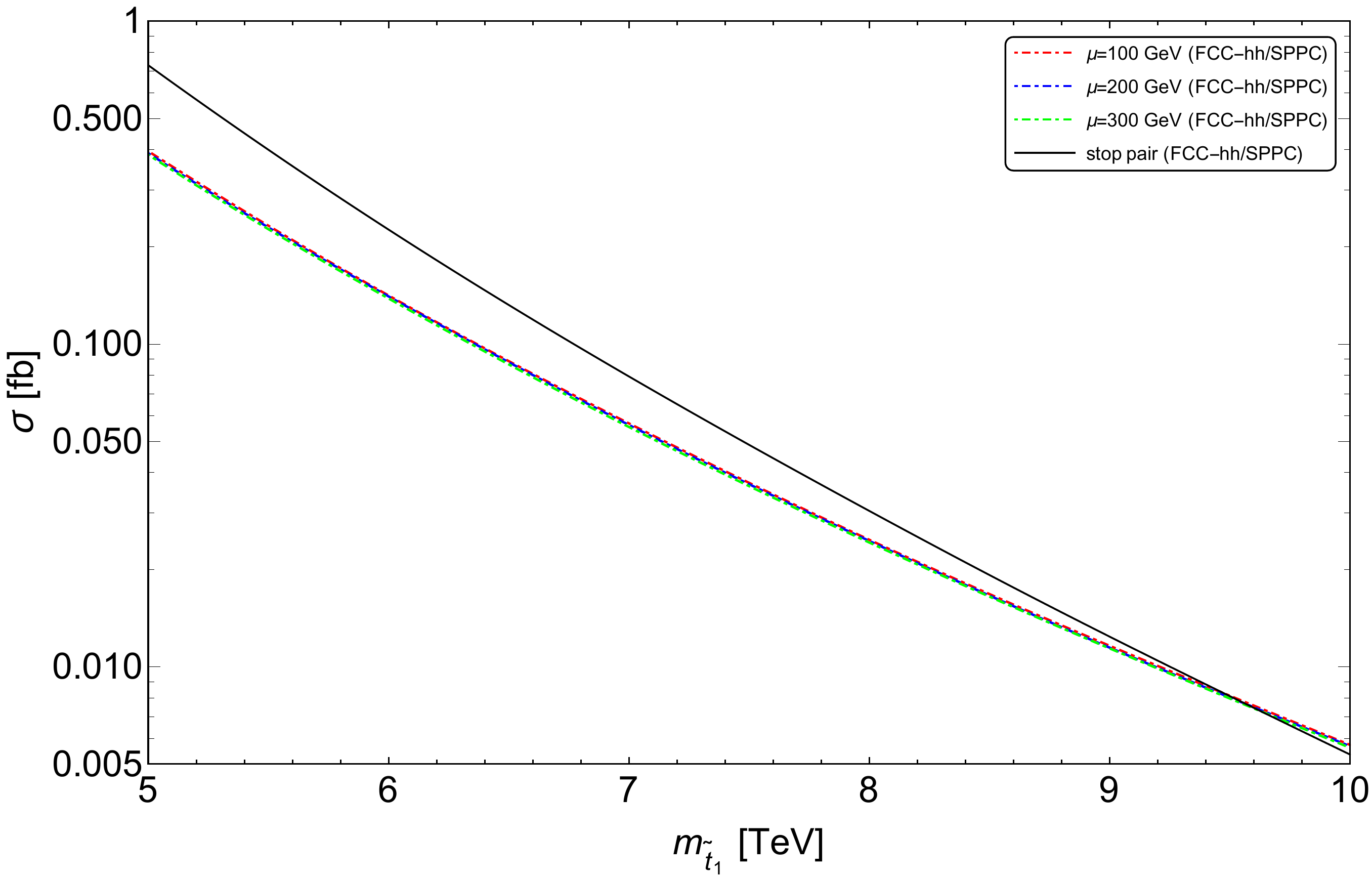}
\caption{The cross sections of stop single production and pair production versus the stop mass at the FCC-hh/SPPC.}
\label{fig:xsforboth}
\end{figure}

In Fig.~\ref{fig:xsforboth} we present the cross sections of stop single production and pair production versus the stop mass. As the stop mass grows, the cross section of single production decreases slower than that of pair production due to a larger phase space. However, taking into account the large SM background, the discovery potential of single stop production still cannot surpass that of pair production. Therefore, the stop pair production has long been considered as the best discovery channel, but the significance of the single stop production via the electroweak interaction, should not be underestimated. Studying the single stop production leads to two implications. On the one hand, the single stop production reveals the electroweak properties of the interaction between stop and neutralinos, which could serve as a complementary channel to its pair production through strong interaction and thus will be an important task for future colliders once the stop is discovered and its mass determined. In principle, the stop pair production can also provide information about the nature of electroweakinos through measuring the kinematic distributions of the products from stop decay, such as the angular distributions. But for a large stop mass, the discriminating power of those observables will be largely reduced due to the boost effects from heavy stop decay~\cite{Wu:2018xiz}. In contrast with stop pair production, the single production process itself is also sensitive to the nature of electroweakinos, and thus may be a complementary search to stop pair production process. On the other hand, the collider signatures of the stop pair production, like $t\bar t$ plus missing transverse energy, can also be present in the signals of other non-supersymmetric models, such as the littlest Higgs Model with T-parity~\cite{Meade:2006dw,Han:2008gy,Yang:2018oek}. Whereas the signatures of the single stop production, such as the mono-b signature, can be helpful in search for the models with top partners~\cite{Goncalves:2017soe}, such as the SUSY and extra dimensional models.

Given that the FCC-hh (Future Circular Collider) and the SPPC (Super proton-proton Collider), proposed projects of hadron-hadron colliders at the center-of-mass energy of 100 TeV, have long been under consideration and extensively studied, the discovery potential for new physics beyond the SM would be increased largely with its high collision energy. In this work, we study the single stop production process in the scenario of natural SUSY at the 100 TeV hadron collider, followed by the mono-top decay channel:
\begin{eqnarray}
&&pp \to \tilde{t}_1 \tilde{\chi}^{-}_{1} \to t \tilde{\chi}^0_{1,2} \tilde{\chi}^-_1.
\label{monotop}
\end{eqnarray}
The top from mono-top channel can further decay into leptonic or full-hadronic final states. In consideration of the large QCD pollution on a hadron collider, we focus on the observability of the leptonic channel. This paper is organized as follows. Sec.~II is the theoretical background of the single stop electroweak production process $pp \to \tilde{t}_1 \tilde{\chi}^{-}_{1}$. Then in Sec.~III, we study the observability of single stop production by performing Monte Carlo simulation of the leptonic mono-top at the 100 TeV hadron collider. Finally we draw our conclusions in Sec.~IV.

\section{single production of stop in a simplified MSSM}

In the minimal supersymmetric standard model (MSSM), the kinetic terms and mass terms of top-squark are given by~\cite{Baer:2006rs}
\begin{alignat}{5}
 \mathcal{L} = (D_\mu \tilde{t}^*_L \, D_\mu\tilde{t}^*_R)\,\left(
\begin{array}{c}
 D^\mu\,\tilde{t}_L \\ D^\mu\,\tilde{t}_R
\end{array}
 \right) - (\tilde{t}^*_L \, \tilde{t}_R^*)\,M_{\tilde{t}}^2\,
\left(
\begin{array}{c}
\tilde{t}_L \\ \tilde{t}_R
\end{array}
 \right),
\label{eq:stop}
\end{alignat}
with the stop mass-squared matrix
  \begin{eqnarray}
  M_{\tilde{t}}^2=
\left(
   \begin{array}{cc}
     m_{\tilde{t}_L}^2 &m_tX_t^{\dag}\\
     m_tX_t& m_{\tilde{t}_R}^2\\
   \end{array}
 \right),
\end{eqnarray}
where
\begin{eqnarray}
&&m_{\tilde{t}_L}^2=m_{\tilde{Q}_{3L}}^2+m_t^2+m_Z^2\left(\frac{1}{2}-\frac{2}{3}\sin^2\theta_W\right)\cos2\beta,\\
&& m_{\tilde{t}_R}^2=m_{\tilde{U}_{3R}}^2+m_t^2+\frac{2}{3}m_Z^2\sin^2\theta_W\cos2\beta,\\
&& X_t = A_t -\mu \cot\beta.
\end{eqnarray}
In the above equations, $A_t$ and $\mu$ are the stop trilinear parameter and the higgsino mass parameter, respectively. The mass eigenstates $\tilde t_1$ and $\tilde t_2$ can be obtained from
\begin{eqnarray}
\begin{pmatrix} \tilde t_1\\ \tilde t_2 \end{pmatrix} &=& \begin{pmatrix} \cos\theta_{\tilde{t}}&& \sin\theta_{\tilde{t}}\\ -\sin\theta_{\tilde{t}} && \cos\theta_{\tilde{t}} \end{pmatrix} \begin{pmatrix} \tilde t_L\\ \tilde t_R \end{pmatrix},
\end{eqnarray}
where $\theta_{\tilde t}$ is the mixing angle between left-handed and right-handed stop.

The electroweakino sector of the MSSM is composed of bino $(\tilde{B})$ , winos $(\tilde{W}^0, \tilde{W}^+, \tilde{W}^-)$ and higgsinos $(\tilde{H}_u^0, \tilde{H}_u^+, \tilde{H}_d^-, \tilde{H}_d^0)$ . The four neutralinos $\tilde\chi_{1,2,3,4}^0$ are mass eigenstates of bino, wino and neutral higgsinos $(\tilde{B}, \tilde{W}, \tilde{H}_d^0, \tilde{H}_u^0)$, whose mass matrix is given by
  \begin{eqnarray}
  \label{mass1}
  M_{\chi^0}=
\left(
   \begin{array}{cccc}
     M_1 &0&-c_\beta s_W m_Z&s_\beta s_W m_Z\\
     0& M_2&c_\beta c_W m_Z&-s_\beta c_W m_Z\\
     -c_\beta s_W m_Z&c_\beta c_W m_Z&0&-\mu\\
     s_\beta s_W m_Z&-s_\beta c_W m_Z&-\mu&0\\
     \end{array}
 \right).
\end{eqnarray}
If $m_Z$ can be neglected, the neutralinos are almost bino-like, wino-like and higgsino-like with masses $M_1, M_2, \mu$. While the two charginos $\tilde\chi_{1,2}^\pm$ are mass eigenstates of charged wino and charged higgsinos $(\tilde{W}^+, \tilde{H}_u^+$, $\tilde{W}^-, \tilde{H}_d^-)$. Similarly, the chargino mass matrix can be written as
 \begin{eqnarray}
  \mathrm{M}_{\chi^{\pm}}=
\left(
   \begin{array}{cc}
     0 &X^T\\
     X& 0\\
     \end{array}
 \right),
\end{eqnarray}
where
\begin{eqnarray}
\label{mass2}
  X=
\left(
   \begin{array}{cc}
     M_2 &\sqrt{2}s_{\beta}m_W\\
     \sqrt{2}c_{\beta}m_W& \mu\\
     \end{array}
 \right).
\end{eqnarray}
If $m_W$ can be neglected, the charginos are almost wino-like and higgsino-like with masses $M_2$ and $\mu$. Therefore, describing the electroweakinos can use just only four electroweakino parameters $M_1, M_2, \mu$ and $\tan\beta$.

\begin{figure}
\centering
\includegraphics[scale=0.6]{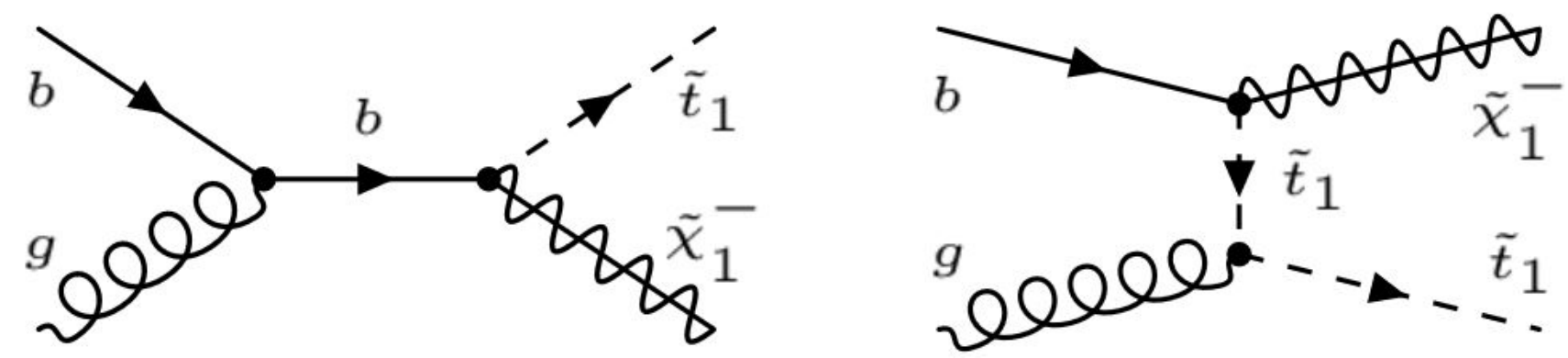}
\caption{Feynman diagrams of the single stop production process $pp \to \tilde{t}_1 \tilde{\chi}^{-}_{1}$ at the partonic level.}
\label{fig:feynmandiagram}
\end{figure}

The relevant couplings between stop and electroweakinos in the mass eigenstates are given by,
\begin{eqnarray}
\mathcal{L}_{\tilde{t}_{1}\bar{t}\tilde{\chi}_{i}^{0}}&=&\bar{t}\left(f^{\tilde{\chi}^{0}}_{L}P_{L}+f^{\tilde{\chi}^{0}}_{R}P_{R}\right)\tilde{\chi}^{0}_{i}\tilde{t}_{1}+\textrm{h.c.}, \label{tt1chi}\\
\mathcal{L}_{\tilde{t}_{1}\bar{b}\tilde{\chi}_{m}^{+}}&=&\bar{b}\left(f^{\tilde{\chi}^{+}}_{L}P_{L}+f^{\tilde{\chi}^{+}}_{R}P_{R}\right)\tilde{\chi}^{+C}_{m}\tilde{t}_{1}+\textrm{h.c.},         \label{bt1chi}
\end{eqnarray}
with $P_{L,R}=\left(1\mp\gamma^{5}\right)/2$, and the coefficients are
\begin{eqnarray}
f^{\tilde{\chi}^{0}}_{L}&=&-\left[\frac{g_{2}}{\sqrt{2}}N_{i2}+\frac{g_{1}}{3\sqrt{2}}N_{i1}\right]\cos\theta_{\tilde{t}}-y_{t}N_{i4}\sin\theta_{\tilde{t}}, \label{fchiL}\nonumber\\
f^{\tilde{\chi}^{0}}_{R}&=&\frac{2\sqrt{2}}{3}g_{1}N^{*}_{i1}\sin\theta_{\tilde{t}}-y_{t}N^{*}_{i4}\cos\theta_{\tilde{t}},
\label{fchiR}\nonumber\\
f^{\tilde{\chi}^{+}}_{L}&=&y_{b}U^{*}_{m2}\cos\theta_{\tilde{t}},
\label{fcharginoL}\nonumber\\
f^{\tilde{\chi}^{+}}_{R}&=&-g_{2}V_{m1}\cos\theta_{\tilde{t}}+y_{t}V_{m2}\sin\theta_{\tilde{t}}.
\label{fcharginoR}
\end{eqnarray}
Here $y_{t}=\sqrt{2}m_{t}/(v\sin\beta)$ is the top quark Yukawa coupling and $y_{b}=\sqrt{2}m_{b}/(v\cos\beta)$ is the bottom quark Yukawa coupling.

For Eq.~\ref{mass1} and Eq.~\ref{mass2}, the mass matrices can be diagonalized by a unitary matrix $N$ and two unitary matrices $U$ and $V$, respectively~\cite{matrix}.
For $M_2 \ll \mu, M_1$, $V_{11}, U_{11} \sim 1$, $V_{12}, U_{12} \sim 0$, $N_{11,13,14}$, $N_{22,23,24} \sim 0$, and $N_{12,21} \sim 1$, the neutralino $\tilde{\chi}^0_{1}$ and the chargino $\tilde{\chi}^\pm_1$ are nearly degenerate winos ($\tilde{W}^\pm$). But for $\mu \ll M_{1,2}$, $V_{11}, U_{11}, N_{11,12,21,22} \sim 0$, $V_{12} \sim \mathop{\rm sgn}(\mu)$, $U_{12} \sim 1$ and $N_{13,14,23}=-N_{24} \sim 1/\sqrt{2}$, the lightest SUSY particles (LSP) are nearly degenerate higgsinos ($\tilde{H}^\pm$). Both of the wino-like and higgsino-like scenarios produce a nearly degenerate mass spectrum of the neutralino and charginos, and hence similar final signatures at a hadron collider. The searches for both scenarios at the LHC have been proposed  ~\cite{giudice-higgisno,wu-higgsino,han-higgsino,park-higgsino,barducci-higgsino,braman-wino,han-wino,ismail-wino}.

\begin{figure}[t]
\centering
\includegraphics[width=8cm,height=8cm]{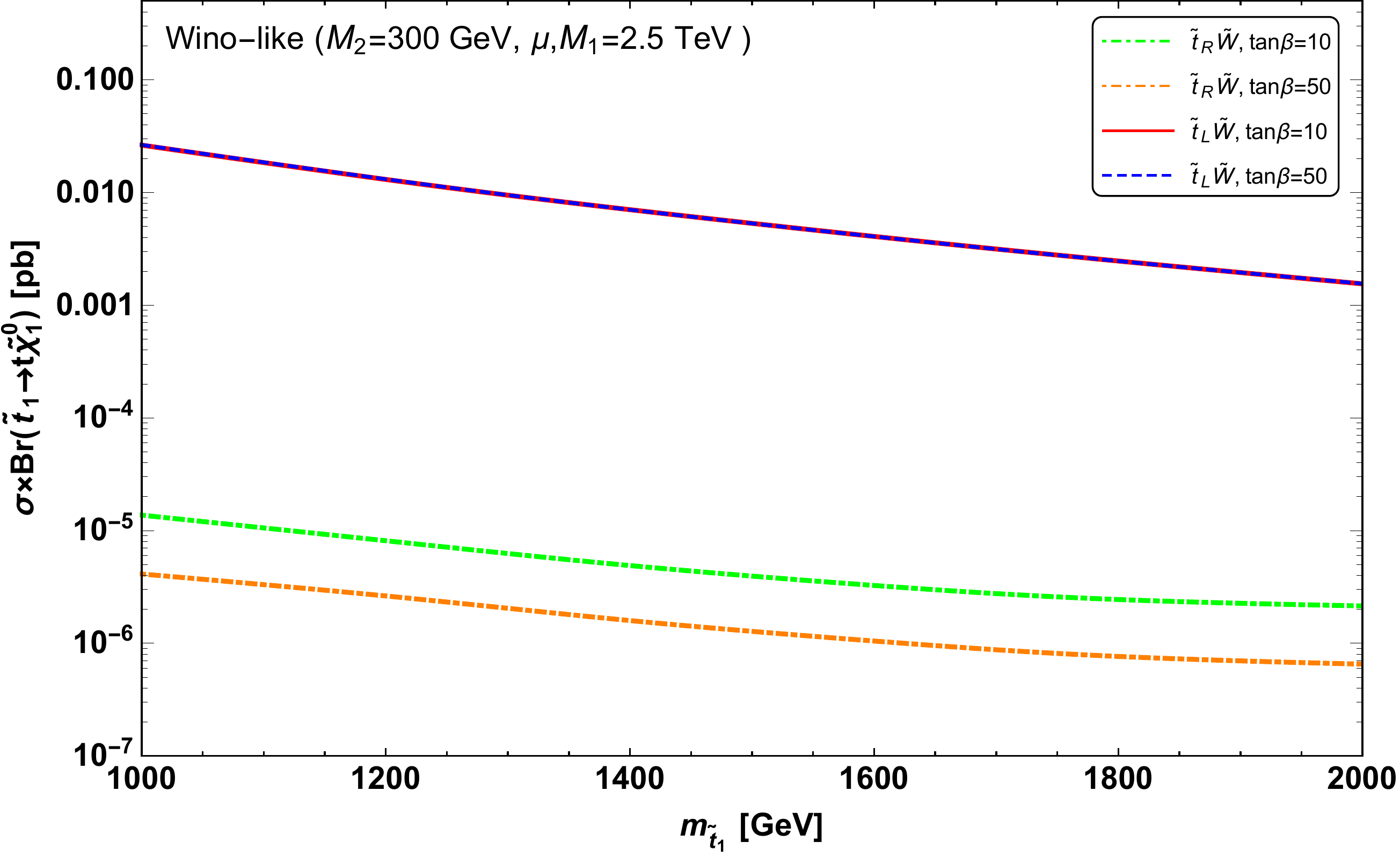}
\includegraphics[width=8cm,height=8cm]{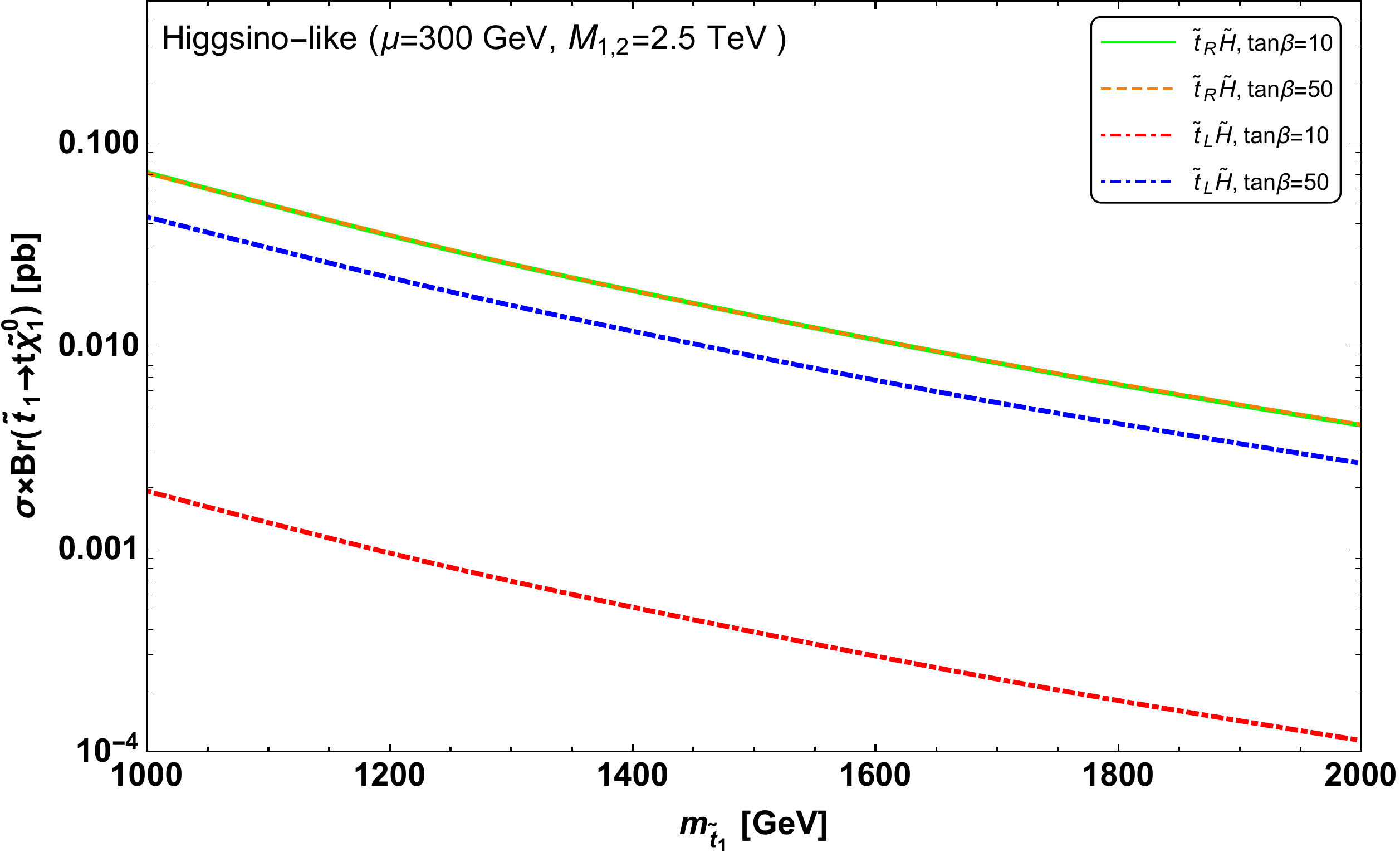}
\caption{The cross sections of the single stop production process $pp \to \tilde{t}_1 \tilde{\chi}^{-}_{1}$ for left- and right-handed stop with $\tan\beta = 10, 50$ at the FCC-hh/SPPC (Left: Wino-like chargino. Right: Higgsino-like chargino). The branching ratios of $\tilde{t}_{R} \to t \tilde{\chi}^{0}_{1,2}$ is assumed to be 50\%. The contribution of the charge-conjugate process of the single stop production $pp \to \tilde{t}^*_{1}\tilde{\chi}^{+}_{1}$ is included.}
\label{fig:xs}
\end{figure}

The partonic process of the single stop production is $g(p_a)b(p_b) \to \tilde{t}_1(p_1) \tilde{\chi}^{-}_{1}(p_2)$, the Feynman diagram of which is shown in FIG.~\ref{fig:feynmandiagram}. 
We present the cross sections of single stop production for left- and right-handed stop under Wino-like and Higgsino-like benchmarks with $\tan\beta=10, 50$ at 100 TeV hadron collider in FIG.~\ref{fig:xs}. In the Wino-like benchmark point, we set $M_{2}=300$ GeV and $\mu, M_{1}=2.5$ TeV, while in the Higgsino-like benchmark point $\mu=300$ GeV and $M_{1,2}=2.5$ TeV. In our simulation, the mass spectrum of sparticles is evaluated by the package SUSYHIT~\cite{SUSYHIT}. Then, we use \textsf{MadGraph5\_aMC@NLO}\,(version 2.6.3)~\cite{Madgraph}  with the NN23LO1 PDF~\cite{Ball:2012cx} to calculate the leading order cross sections of the stop single production process $pp \to \tilde{t}_1\tilde{\chi}^+_1$. The QCD corrections at next-to-leading order are included by applying a K-factor of 1.4~\cite{Jin:2003ez,Jin:2002nu,electroweak}. From FIG.~\ref{fig:xs}, we can learn that for wino-like $\tilde{\chi}^\pm_1$, because of the gauge interactions, the cross section of the left-handed stop $\tilde{t}_L\tilde{W}^-$ production is larger than that of the right-handed stop $\tilde{t}_R$. However, for higgsino-like $\tilde{\chi}^\pm_1$, the cross section of the right-handed stop $\tilde{t}_R$ single production is much larger. In addition, the cross section of $\tilde{t}_R\tilde{H}^-$ is insensitive of $\tan{\beta}$, but $\tilde{t}_L\tilde{H}^-$ is not. This is because the coupling of $\tilde{t}_L$ with $\tilde{\chi}^\pm_1$ is dominated by the bottom Yukawa coupling and can be enhanced as the value of $\tan{\beta}$ increases.

\section{Observability of mono-top signature at the FCC-hh/SPPC}
We use $\tilde t_R\tilde H^-$ production to study the observability of the leptonic mono-top signature for the single stop production at the 100\,TeV hadron collider. In our Monte Carlo simulation, we set $\mu \ll M_{1,2}$, $m_{\tilde U_{3R}} \ll m_{\tilde Q_{3L}}$ and $\tan\beta = 50$, thus the electroweakinos are higgsino-like and the stop is right-handed. It should be noted that the branching ratios of $\tilde{t}_{R} \to t \tilde{\chi}^{0}_{1,2}$ is about 50\%~\cite{Duan:2016vpp}. In the following study, we focus on a simplified MSSM framework where the higgsinos and right-handed stop are the only sparticles. We generate the parton-level signal and background events with \textsf{MadGraph5\_aMC@NLO}\,(version 2.6.3). Then within the framework of \textsf{CheckMATE2}\,(version 2.0.26)~\cite{checkmate}, we use \textsf{Pythia-8.2}~\cite{pythia} and \textsf{Delphes-3.4.1}\,(using ATLAS Delphes card)~\cite{delphes} to implement parton shower and detector simulation, respectively. Given that the mass splittings $m_{\tilde \chi^\pm_1}-m_{\tilde \chi^0_{1}}$ and $m_{\tilde \chi^0_2}-m_{\tilde \chi^0_{1}}$ are small, thus the chargino $\tilde \chi^\pm_1$ and the neutralino $\tilde \chi^0_{2}$, like the LSP $\tilde \chi^0_{1}$, are also treated as missing transverse energy $\slashed E_T$ in our simulation. We adopt the b-jet tagging efficiency as 80\%~\cite{cms-b} with MV2c20 algorithm~\cite{mv2c20} and cluster the jets by the anti-$k_t$ algorithm with the cone radius $\Delta R=0.4$~\cite{anti-kt}.

\begin{figure}[t]
\centering
\begin{minipage}{0.48\linewidth}
  \centerline{\includegraphics[width=8cm,height=7cm]{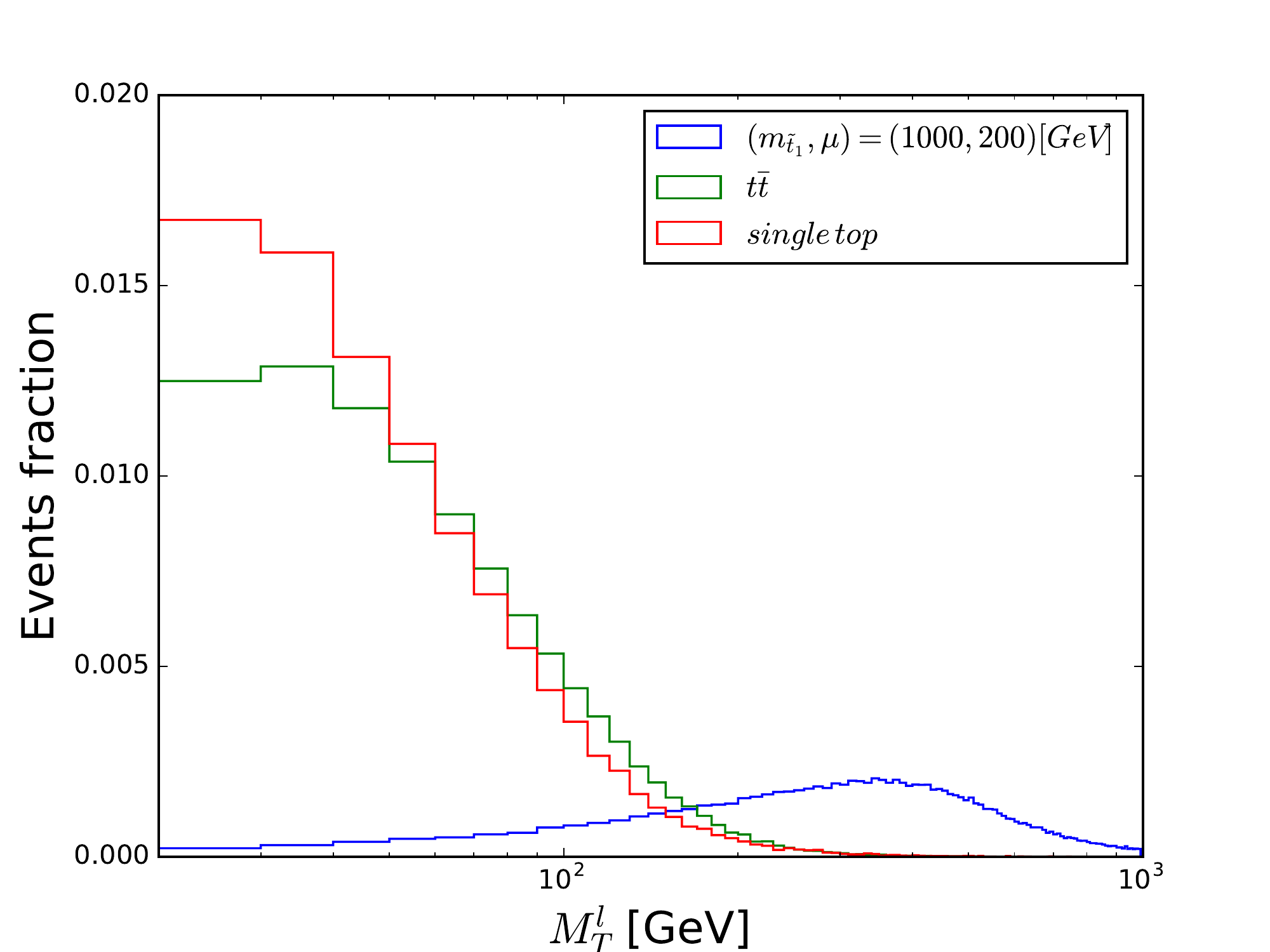}}
  \centerline{(a)}
\end{minipage}
\hfill
\begin{minipage}{0.48\linewidth}
  \centerline{\includegraphics[width=8cm,height=7cm]{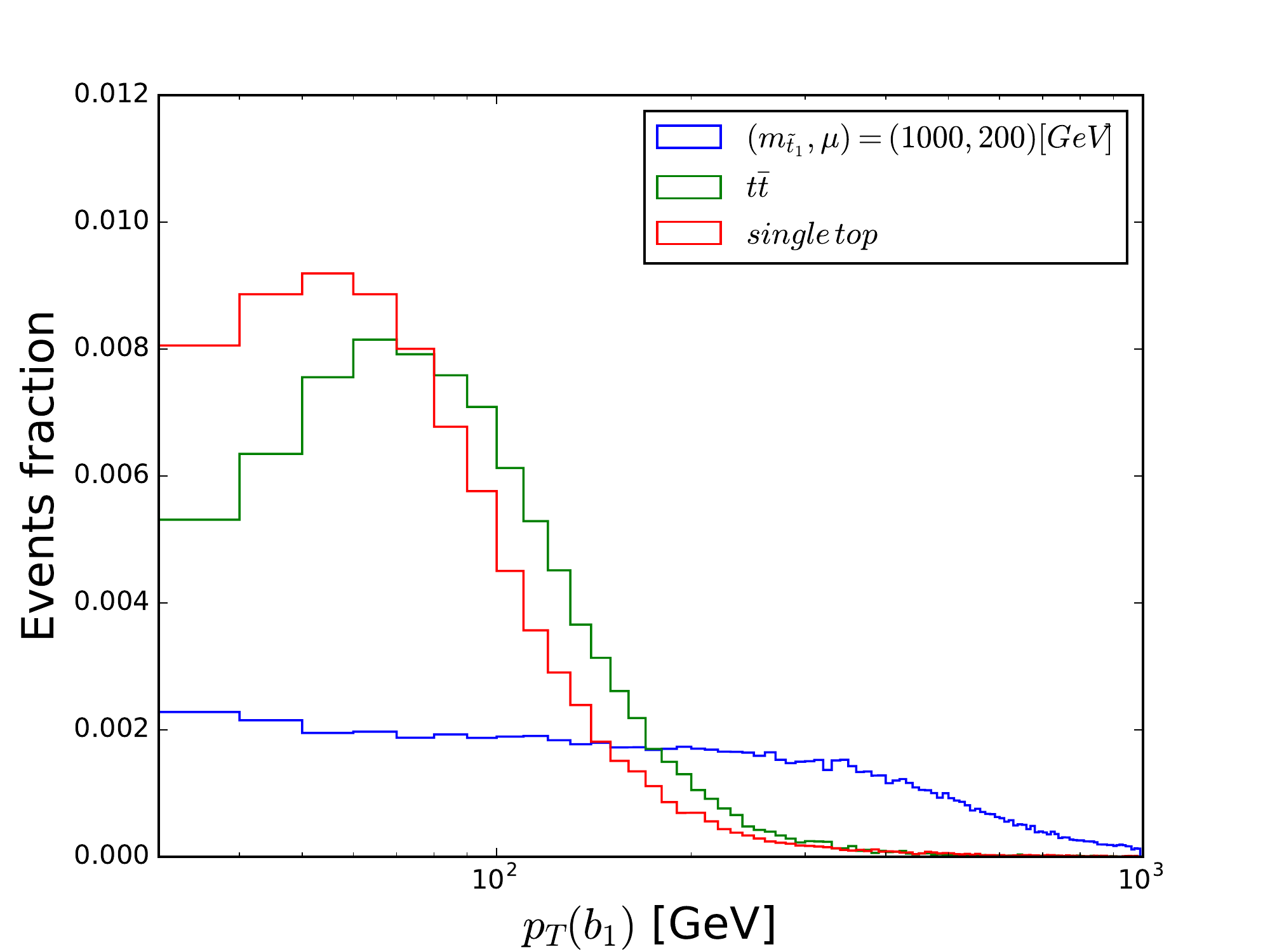}}
  \centerline{(b)}
\end{minipage}
\hfill
\begin{minipage}{0.48\linewidth}
  \centerline{\includegraphics[width=8cm,height=7cm]{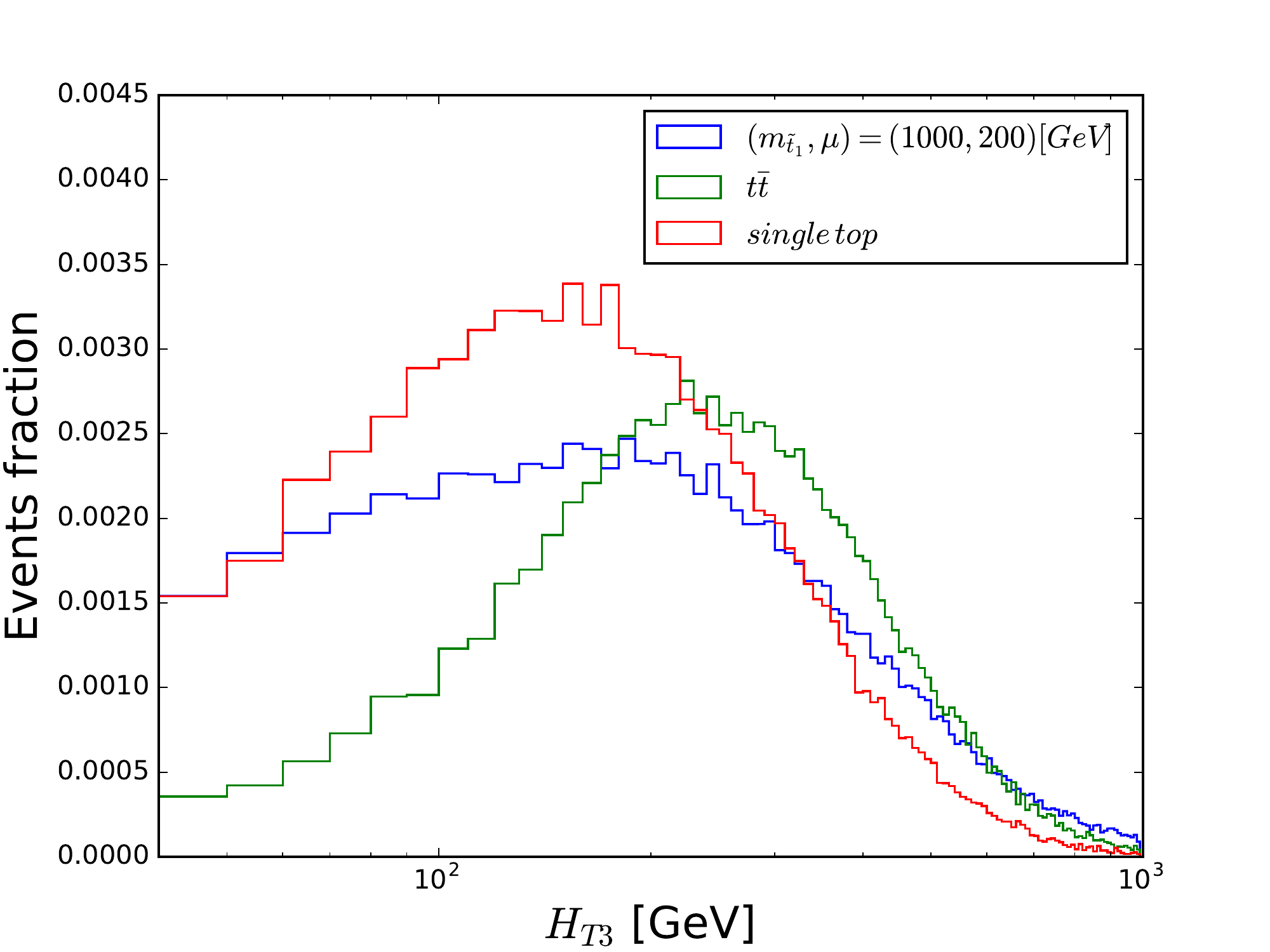}}
  \centerline{(c)}
\end{minipage}
\hfill
\begin{minipage}{0.48\linewidth}
  \centerline{\includegraphics[width=8cm,height=7cm]{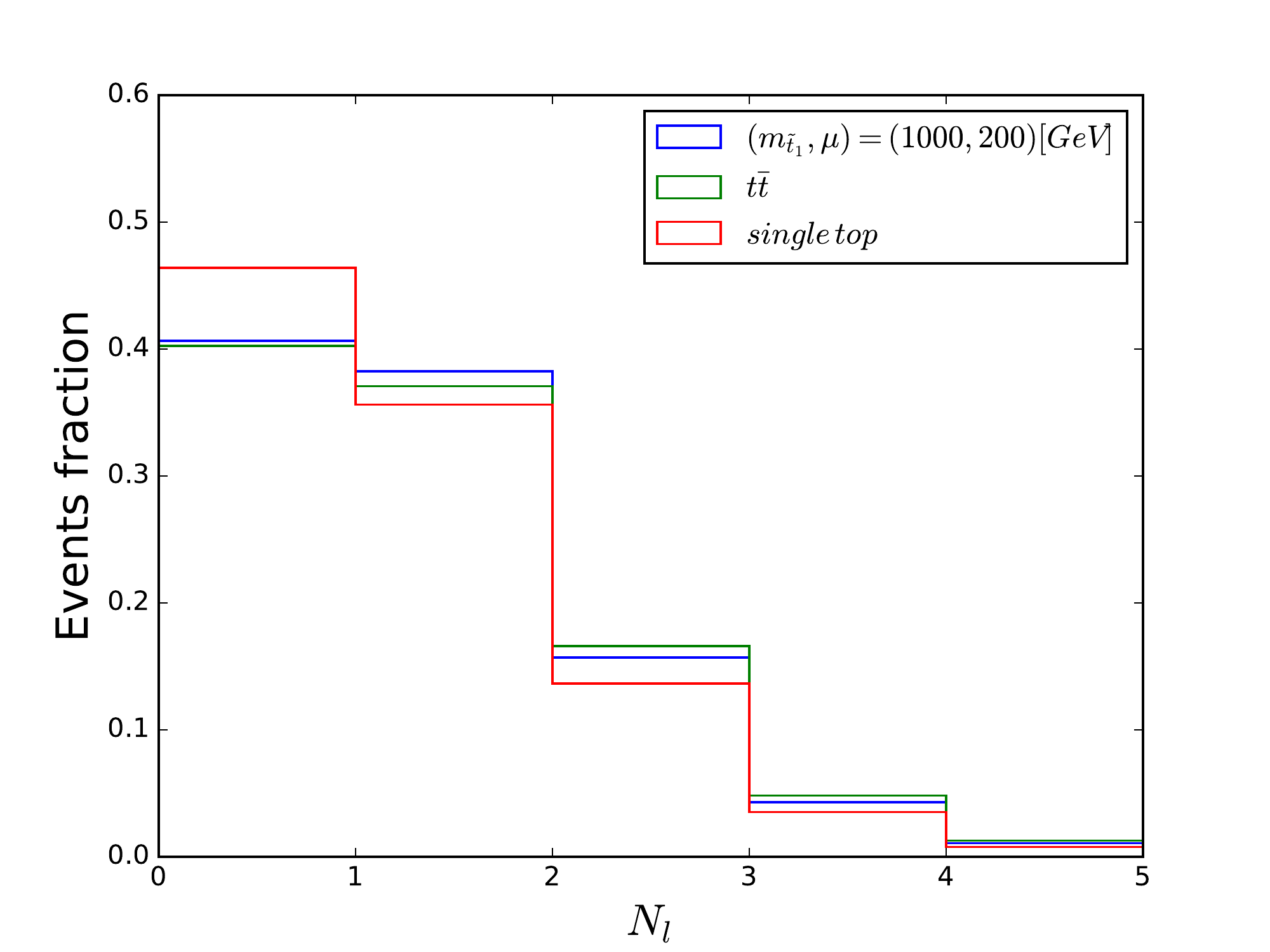}}
  \centerline{(d)}
\end{minipage}
\caption{The normalized distributions of $M^l_T$, $p_{T}(b_{1})$, $H_{T3}$ and $N_l$ for the mono-top channel and background events at the FCC-hh/SPPC. The benchmark point is $m_{\tilde{t}_1}=1000$ GeV and $\mu=200$ GeV.}
\label{fig:distribution_monotop}
\end{figure}

For the leptonic mono-top channel decay $\tilde{t}_1 \to t \tilde{\chi}^0_{1,2} \to (b \ell^+ \nu) \tilde{\chi}^0_{1,2}$, the largest SM background comes from the $pp\rightarrow t\bar t$ production followed by semi- and di-leptonic decay, 
because of the undetected leptons and the limited jet energy resolution that lead to large $\slashed E_T$. The single top production including $pp\rightarrow tj$, $tb$ and $tW$ can also fake the signal due to the missing leptons. Backgrounds from the diboson production, such as $WW$, $WZ$ and $ZZ$, will not be considered owing to their relatively small cross sections. In order to enhance the signal, some kinematic cuts are applied to suppress the background. The transverse mass of the lepton plus missing energy $M^l_T$ is used since the final lepton and missing energy of the backgrounds come from $W$ decay, leading to an end-point to separate from the signal events~\cite{pdg}. The signal has one hard b-jet in the final state, thus the transverse momentum of the leading b-jet $p_T(b_1)$ can be used to suppress the background. As the signal has fewer hard jets in the final state, $H_{T3}$, the scalar sum of the transverse momentum of jets excluding the leading and subleading ones, can suppress the $t\bar t$ background effectively~\cite{HT3}.

In FIG.~\ref{fig:distribution_monotop}, we present the normalized distributions of $M^l_T$, $p_T(b_1)$, $H_{T3}$ and $N_l$ for the signal and backgrounds at the FCC-hh/SPPC. The benchmark point is $m_{\tilde{t}_1}=1000$ GeV and $\mu=200$ GeV. From the curves of $M^l_T$ and $p_T(b_1)$ in FIG.~\ref{fig:distribution_monotop}\,(a) and FIG.~\ref{fig:distribution_monotop}\,(b), we can find that the ones for signal events tend to be more flat and smooth, while the ones for the SM background tend to distribute around the small $M^l_T$ and $p_T(b_1)$, which are well separated from the signal. In FIG.~\ref{fig:distribution_monotop}\,(c), one can see that $H_{T3}$ of signal events tends to be smaller than that of background events, as we infer in the above analysis. We also show the distributions of $N_{l}$ for signal and backgrounds in FIG.~\ref{fig:distribution_monotop}\,(d), which is the number of final leptons, from which one can find that more leptons ($N_{l}>1$) tend to be found in the signal events.

\begin{table}
\centering
\begin{tabular}{|c|c|c|c|}
\hline\hline
Cut & Signal & \multicolumn{2}{c|}{Background}\\
\hline
$m(\tilde{t}_{1}, \mu)$ [GeV] &  $(1000, 200)$  & $t\bar t$ & single top \\
\hline
$N_l \ge 1$ & 442.14 & $1.44 \cdot 10^7$ & $2.94 \cdot 10^6$ \\
\hline
$N_b \ge 1$ & 406.62 & $1.34 \cdot 10^7$& $2.61 \cdot 10^6$ \\
\hline
$p_T(b_1) > 250$ GeV &216.08 &$5.66 \cdot 10^5$& $1.07 \cdot 10^5$ \\
\hline
$\slashed E_T > 850$ GeV  &32.91 & $9.08 \cdot 10^2$& $1.17 \cdot 10^3$ \\
\hline
$M^l_T > 900$ GeV &28.33 & $5.96 \cdot 10^2$& $9.36 \cdot 10^2$  \\
\hline
$H_{T3} < 300$ GeV &3.56 & 12 & 15.5 \\
\hline
$\Delta \phi(j,\slashed E_T) > 0.6 $ &1.93 & 2& 3.6 \\
\hline
\end{tabular}
\caption{A cut flow analysis of the cross sections for the mono-top channel and backgrounds at the FCC-hh/SPPC. The cross sections are shown in unit of fb.}
\label{tab:cutflow1}
\end{table}

\begin{figure}[b]
\centering
\includegraphics[scale=0.54]{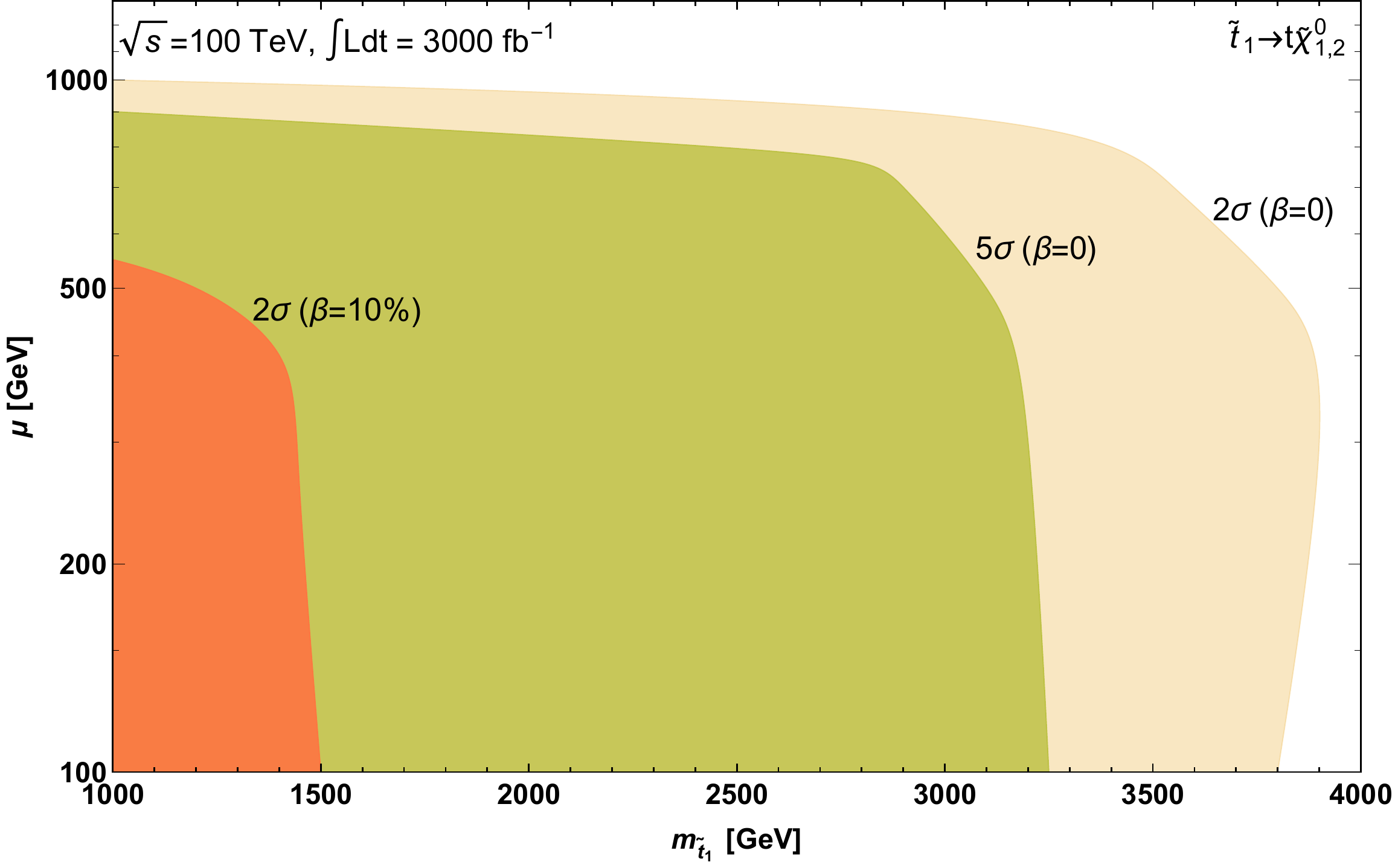}
\caption{The statistical significance $\alpha=S/\sqrt{B+(\beta B)^{2}}$ of the mono-top channel on the plane of stop mass $m_{\tilde{t}_{1}}$ versus the higgsino mass parameter $\mu$ at the FCC-hh/SPPC. $\beta$ is taken as 0 and 10\% as an estimate for the impact of systematic uncertainty.}
\label{fig:contour}
\end{figure}
According to above distributions and analysis, the following cuts are applied:
\begin{itemize}
  \item At least one lepton is required.
  \item At least one b-jet with $p_T(b_1) > 250$ GeV is required.
  \item We define five signal regions according to $(\slashed E_T, M^l_T)$ cuts: (750, 800), (800, 850), (850, 900), (900,950), (950, 1000). They can well separate the backgrounds and signal.
  \item We require $H_{T3} < 300$ GeV to further suppress the top pair background events.
  \item A minimum azimuthal angle between $\slashed E_T$ and each of the jets $\Delta\phi(j,\slashed{E}_{T})>0.6$ is required to reduce the multi-jet events from $t\bar{t}$ background.
\end{itemize}
The cutflow of signal and background at every step of the above cuts is shown in TABLE~\ref{tab:cutflow1}, from which we can see that after these cuts, the background can be suppressed significantly with a relatively large amount of signal events surviving. In order to estimate the signal significance ($\alpha$), we adopt the formula $\alpha=S/\sqrt{B+(\beta B)^{2}}$ in which $S$ and $B$ stand for number of signal and background events after our cuts, respectively. $\beta$ is the systematic error. In FIG.~\ref{fig:contour}, we display the contour in the plane of the higgsino mass parameter $\mu$ and stop mass $m_{\tilde{t}_{1}}$, at the statistical significance of $2\sigma$ and $5\sigma$ with the center-of-mass energy of 100\,TeV and integrated luminosity of $3000\,\text{fb}^{-1}$, from which we find that, for $\beta=0$, top squark with mass up to 3.25\,TeV can be probed at $5\sigma$ level through the single stop production followed by a mono-top leptonic decay channel. The exclusion limits for stop mass and higgsino mass parameter can be reached at about 3.9\,TeV and 1\,TeV, respectively. As a comparison, the current LHC search for stop, based on analysis of 139 $\text{fb}^{-1}$ of 13 TeV collision data~\cite{Aad:2020sgw}, has excluded the stop mass in the range 400$\sim$1250 GeV at 95\% CL depending on the LSP mass. And to estimate the impact of systematic uncertainty, we assume $\beta$ to be 10\% and find the exclusion limit for stop mass to be about 1.5 TeV. Note that a search for the hadronic mono-top channel has been performed at the 14 TeV LHC~\cite{Fuks:2014lva}, from which we can infer that the present search for the leptonic channel is comparable to the hadronic one, but further study is needed. It should be mentioned that the statistical significance would get worse when considering systematic uncertainties, the determination of which due to high pile-up in the future must be revisited with the real performance of upgraded detectors. In addition, our results may be improved by using some advanced analysis approaches, such as the recently proposed machine-learning methods for sensitivity enhancement in searching for sparticles at the LHC ~\cite{Albertsson:2018maf,Abdughani:2019wuv,Ren:2017ymm,Caron:2016hib}.

It should be noted that the best discovery channel for stop may still be the pair production. The single production, however, provides an instructive way to learn more about the SUSY particles and to explore more specific models.
As we concluded in the Sec.~II, for higgsino-like $\tilde{\chi}^\pm_1$, the single production cross section of the right-handed stop is much larger than that of the left-handed stop and independent of $\tan{\beta}$.
In addition, the Focus Point SUSY model usually has the LSP chargino as higgsino-like, while the AMSB SUSY model prefers a wino-like one. The single production has a larger cross section in the former case than in the latter one with the same stop and chargino masses~\cite{wu-1,Duan:2016vpp}. It should be emphasized again that the study of electroweak production of the stop is meaningful with the stop either discovered or even not observed. Furthermore, future precision measurement on the cross section of stop single production can be used to investigate the nature of stop and electroweakinos, which is also helpful in identifying different SUSY models. For example of the stop mass at 2 TeV in the higgsino scenario with tan$\beta=10$, the single production cross section of $\sim10^{-2}$ pb corresponds to a right-handed stop, while the cross section of $\sim10^{-4}$ pb corresponds to a left-handed one, as can be seen from FIG.~\ref{fig:xs}. By using the effective field theory (EFT) techniques to study the little hierarchy problem of the SM Higgs sector, it is found~\cite{Bar-Shalom:2014taa} that heavy new physics theories that can restore naturalness in the effective action at e.g., $\Lambda \simeq 5\sim10$ TeV, should include one or more singlet or triplet heavy bosons or else a singlet, doublet or triplet fermions, all having typical masses larger than $\Lambda$. The precision electroweak data and the recently measured Higgs signals impose on the EFT-naturalness, which requires heavy scalar singlets and/or heavy fermions (singlets, doublets or triplets) are more likely to play a role in softening the fine-tuning in the SM Higgs sector, if the scale of the new heavy physics is below $\simeq 10$ TeV.

\section{CONCLUSION}
In this work, we studied the mono-top decay channel of single stop production in a simplified MSSM framework where the higgsinos and stops are the only sparticles at the future hadron colliders FCC-hh and SPPC. The single stop production leads to different signals from traditional stop pair production, which ends up with final states of $t\bar{t}$ plus missing energy. We performed Monte Carlo simulation to study the observability of the mono-top channel and found that through single stop production followed by the leptonic mono-top channel, we can probe the stop mass up to 3.25\,TeV at $5\sigma$ statistical significance at the 100\,TeV hadron collider with integrated luminosity of 3000\,$\text{fb}^{-1}$, while the exclusion limits for stop mass and higgsino mass parameter $\mu$ are 3.9\,TeV and 1\,TeV, respectively. Including the impact of systematic uncertainty (10\% as an example), the exclusion limit for stop mass can be reached at about 1.5 TeV.

\section*{Acknowledgement}
This work was supported by the National Natural Science Foundation of China (NNSFC) under grant Nos. 11847208 and 11705093, and by the Jiangsu Planned Projects for Postdoctoral Research Funds, Grant No.2019K197.

\end{document}